%% file: template.tex
\documentclass[
aps,%
12pt,%
final,%
notitlepage,%
oneside,%
onecolumn,%
nobibnotes,%
nofootinbib,%
superscriptaddress,%
noshowpacs,%
centertags]%
{revtex4-1}
\usepackage{float}
\usepackage{graphicx}
\usepackage{natbib}
\newcommand{\etal}{\mbox{\rm{et al.}~~}}
\newcommand{\kms}{\mbox{km\,s$^{-1}$}}
\begin{document}
%



\title{
OORT CLOUD BOMBARDMENT BY DARK MATTER}


\author{Jeremy Mould }
\email{jmould@swin.edu.au}
\affiliation
{Swinburne University, John Street, Hawthorn, Victoria 3122, Australia}

\affiliation
{Centre of Excellence for Dark Matter Particle Physics, The University of Melbourne, Grattan Street, Parkville, Victoria, 3010, Australia.}






\begin{abstract}
The realization that primordial black holes (PBHs) might be  some fraction of the dark matter 
begged the question, how often do PBHs enter the solar system? For a Neptune radius solar system the answer is, rarely.
For an Oort cloud sized system the answer is different. Simulations of bombardment of the Oort cloud by dark matter
suggest that dislodgement of protocomets and their entry into the inner solar system can match the observed frequency of comets, if that PBH fraction is high 
enough.
Comets were traditionally considered as messengers, usually omens.
After 50 years of puzzlement regarding dark matter, we need a hint
from the dark universe about the size and nature of dark matter particles.
\end{abstract}



\maketitle

\input comet.tex

%% file: comet.tex
\section{Introduction}
Besides nearby planets, comets were the first recorded periodic astronomical phenomena. Could they also
be telling us about the first component of our Galaxy to form, its dark halo?
The tidal force from the Galactic disk is believed to be the dominant external force in the evolution of the bodies in the Oort cloud (Dones \etal 2004). 
The vertical component (i.e., perpendicular to the Galactic plane) of the Galactic tide is thought to play the most important role in the formation of the Oort 
cloud and the production of long-period comets from it (e.g., 
Harrington 1985, Heisler \& Tremaine 1986),
but a radial component is also included by Higuchi (2007). Collins \& Sari (2010) note that both passing stars and tidal torques are factors to be taken into account.

This may not be the only force impinging on the Oort cloud, however.
If dark matter consists of macroscopic particles as opposed to WIMPs (Weakly Interacting Massive Particles) or axions, 
the effects may be observable 
in the form of comets. Here we focus on lunar mass particles, nominally 10$^{-7}
$ M$_\odot$. These might be primordial black holes (PBH, 
Murai, Sakurai \& Takahashi 2025); they might be axion miniclusters (Fairbairn \etal 2025),
or they might be free floating moons (FFM). Microlensing experiments are unable
to tell the difference (K\"uhnel 2025). Here one can be  agnostic as to which dark 
matter (DM) particles form the local density of
0.01 M$_\odot$ pc$^{-3}$. But it should be noted that it is not tenable that the
 rest of the DM in the universe is 
FFMs, as they are baryonic, and the baryonic fraction of matter is 0.0224/0.12 (Planck collaboration 2020).
Like other DM candidates, both PBHs and FFMs lack an  understanding of the exact formation mechanism.

Other signs of DM activity in the solar system have been suggested. De Rocco (2025)
notes that DM in the form of macroscopic composites is largely unconstrained at 
masses of 10$^{11}$ -- 10$^{17}$g. In this mass range, DM may collide with planetary bodies, depositing energy and leaving dramatic surface features that remain
 detectable on geological timescales. He suggests that 
Ganymede, the largest Jovian moon, provides a prime target to search for DM impacts due to its differentiated composition and Gyr-old surface.
There are also searches for microscopic dark matter, such as WIMP capture in the Sun.
Stothers (1984) was first to connect DM and comets,  after the appearance of heavy halo models
of the Galaxy (e.g. Caldwell \& Ostriker 1981). Brown, He \& Unwin (2025)
have explored the effect of PBH on exoplanet systems.

The collision rate for objects incident on the solar system is $\sigma$ v n , where $\sigma$
is the geometric cross section for the solar system, v is their velocity 
and n is their number density. The ratio of dark matter rate to stellar rate is (n$_{DM}$/n$_*)~\times$ (v$_{DM}$/v$_*$) which equals ($\rho_{DM}$/m$_{DM})~\times$~v$_{DM}$ divided by
($\rho_*$/m$_*)~\times$ v$_*$, where $\rho$ is density and m is mass. For m$_{DM}$ = 10$^{-7}$ M$_\odot$, the rate ratio is 10$^7$ 0.01/0.1 220/30 for old disk stars 
that dominate the solar neighbourhood. So the DM bombardment rate is ten million times larger than the stellar rate. 
In fact, the conventional theory is not that stars perturb the Oort cloud. It is that the 
tidal field of the Galaxy perturbs the Oort cloud. 

The nature of the DM remains one of the outstanding questions of 21st century astrophysics.
Microlensing observations with the Rubin telescope could be prioritized to discover subsolar mass objects
(Romao, Croon \& Godines 2025).
Strong limits on fundamental particle DM are being set by physicists (Aalbers \etal 2023,
Barberio \etal 2025). These may nevertheless be an important component of the DM,
but they do not dislodge protocomets. 

In the next section a toy model is developed to examine the effectiveness of lunar mass DM.
In $\S$II results of these simulations are presented. They are discussed in $\S$III. The
conclusion is that within the mass range discussed macroscopic DM is a plausible contributor
to the rate of comet production.
\section{A toy model}
 The cumulative number of protocomets in the Oort cloud is given as N $\propto$ 
D$^{-\alpha}$, for D $\sim$ 2.8 km, with $\alpha$ = 3.6 for larger D $>$ 2.8 km, and $\alpha$
 = 0.5 for smaller objects (Wajer \etal 2024). 
 The Oort cloud contains $\sim$ 3.8 $\times$ 10$^8$ comets with  D $>$ 10 km 
 (Nesvorn\'y 2018).
Extending this to  decimeter size (Vida \etal 2023) 
we obtain N = 1.5 $\times$ 10$^{14}$, compared with
 10$^{12}$ -- 10$^{13}$ (Wiegert \& Tremaine 1998). 
The Oort cloud is taken to be 
spherical
, which is appropriate to its outer parts. 
The number of comets in  the simulation is 250,000, compared with our adopted
N = 4 $\times$ 10$^{13}$
 and so the fraction of 4$\pi$ sr in the simulation is given by $\pi \theta^2/4\pi$ = 2.5 $\times$ 10$^5$/N,
where  $\theta$ is the half angle.

The adopted distribution radius r from the sun is 
(Dehnen 1993) a number density of
$$ n \propto \frac{a}{r^2 (r+a)^2}      \eqno(1)$$
where a = 10$^5$ AU and the rate of intrusion by lunar mass objects 
can be calculated from the number density 
and the object velocity, taken to be, without loss of generality,
 in the yz plane, where z is the axis from the Sun at the origin to the
Oort Cloud protocomet. Intercometary forces are neglected, 
again without much loss of generality. Distributions by  Duncan \etal (1987)
and Dones \etal (2004) are compared in Figure 1.
Kaib \etal (2011) note that the radius of the Oort cloud, a,
may have varied over time with the migration of the Sun
in its radial distance from the Galactic Centre.
\begin{figure}
	\includegraphics[width=\textwidth]{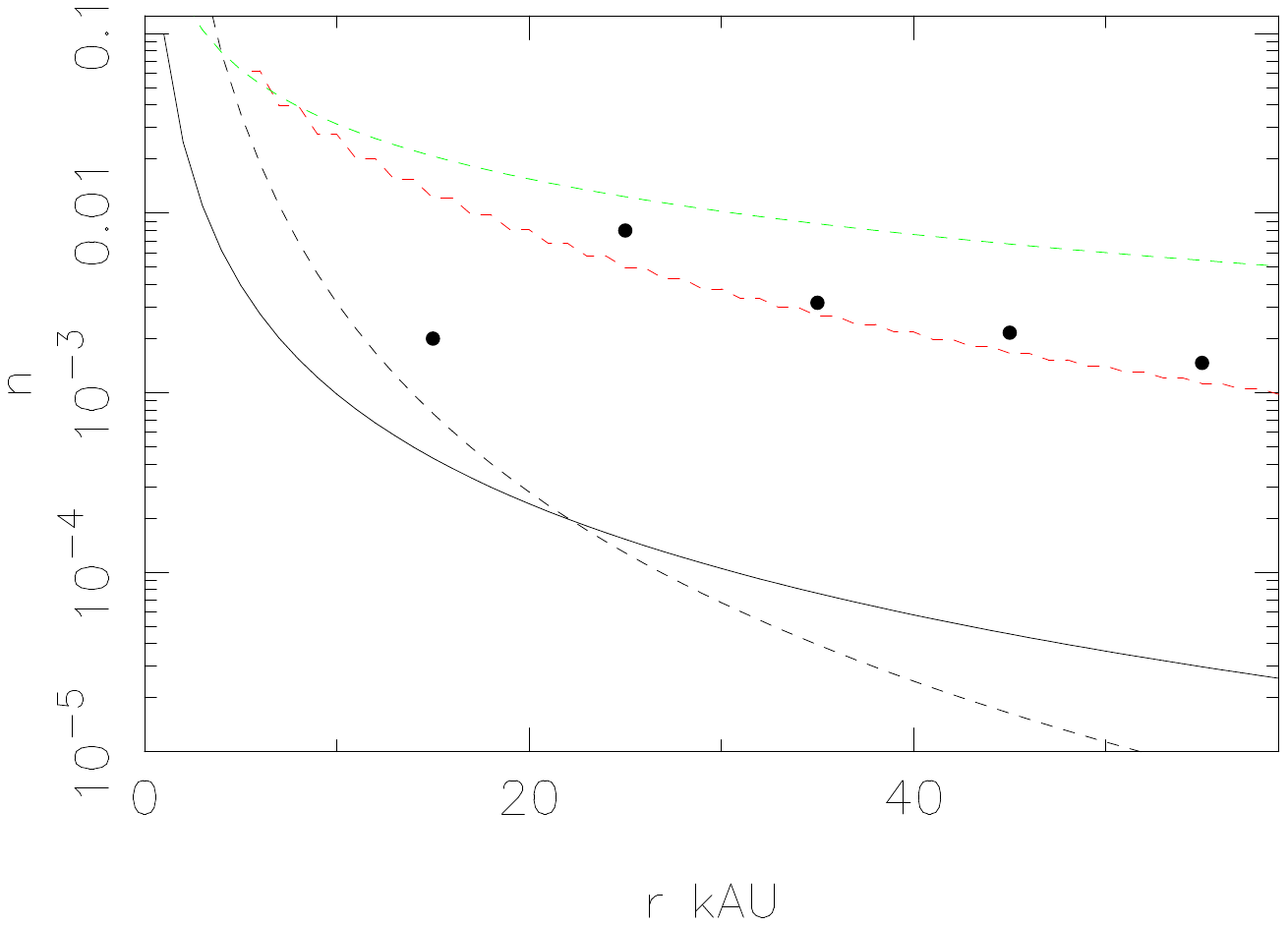}
	\caption{The Dehnen (1993) distribution in the solid line.
	Other authors suggest a radial power law with index --3.5,
	and this is  shown by the dashed line.
	Brasser \etal (2006) finds that the median  distance falls off
	as $\surd$n.  Two variants of this density law are shown as the red and green dashed lines respectively. The distribution adopted by Kaib \& Quinn (2008) is represented
	by the solid symbols (black dots), based on the cumulative
	distribution they provide.}
\end{figure}
There are other density distributions in the literature and these are compared
in Figure 1.
The velocity distribution of the perturbers has two components, that due
to the Sun's galactic orbit, 220 \kms in the yz plane and the galactic halo's
110 \kms velocity dispersion with an isotropic distribution.
Solar system units are adopted with M = v$_{30}^2$ r with masses in solar units,
 v$_{30}$ in Earth orbital velocity units, r in AU, and G = 1. 

The impact parameter, b,  of the incoming object is a crucial parameter. Unless a maximum value is adopted, beyond which the possibility of collision approaches zero, the computing time diverges, as all values of b are plausible out to the mean separation radius of protocomets.  I therefore
assume b lies with uniform probability within a sphere  
	of radius b$_{max}~\approx$ M AU, where M
is the mass of the object in M$_\odot$ units. 
For an impact velocity, v, this gives a momentum transfer,
when the interaction with the protocomet is complete, of,

$$\Delta p = 2 G M m / b v        \eqno(2)$$

with m the mass of the perturbed particle, whose velocity is 
parallel to the vector from its original position to the impact parameter
position. This is the impulse approximation,
derived in full by Rickman (1976), the
principal assumption being a high value of $v$.
We then follow the orbits of 250,000 such particles with time intervals
of 0.10 
to 0.3 yr. 
The appropriate time interval was monitored by noting when comets entering the inner solar system were double counted. The integration scheme was simply {\bf $\delta$v} = {\bf a}~dt, where {\bf a} is the acceleration, followed by {\bf $\delta$r} = {\bf v}~dt, where {\bf v} is the velocity.
Simulations
with different DM masses and other parameters are recorded
in Table I. Any protocomet reaching the inner solar system with radius 300 AU
is noted. This choice of radius is on the large size, but the simulations
are limited by small number statistics, and a markedly smaller radius
would be prohibitive in computer time and not necessary for this first exploration
of the hypothesis.  

With these initial conditions we are neglecting the perturbers that arrive
with impact parameters larger than b$_{max}$, and also oversupplying impacts by
a factor 4(a/b$_{max})^2$/N, where N is the number of protocomets in the Oort cloud. 
The arrival rate of DM particles at the Oort cloud\footnote
{
As a spherical shell the DM traverses
it both on the windward and the leeward side. There is also a DM velocity
dispersion of 110 \kms (Freeman 1985).}
is rate$_1$ $\approx$ 4$\pi \rho a^2 v/m$, where $\rho$ is the local density of
DM, 0.01 M$_\odot pc^{-3}$ (de Salas \& Widmark 2021) and v is the solar
velocity of 220 \kms (Binney 2011). 
\begin{figure}
\hspace{-2.0cm}
\includegraphics[width=0.75\textwidth]{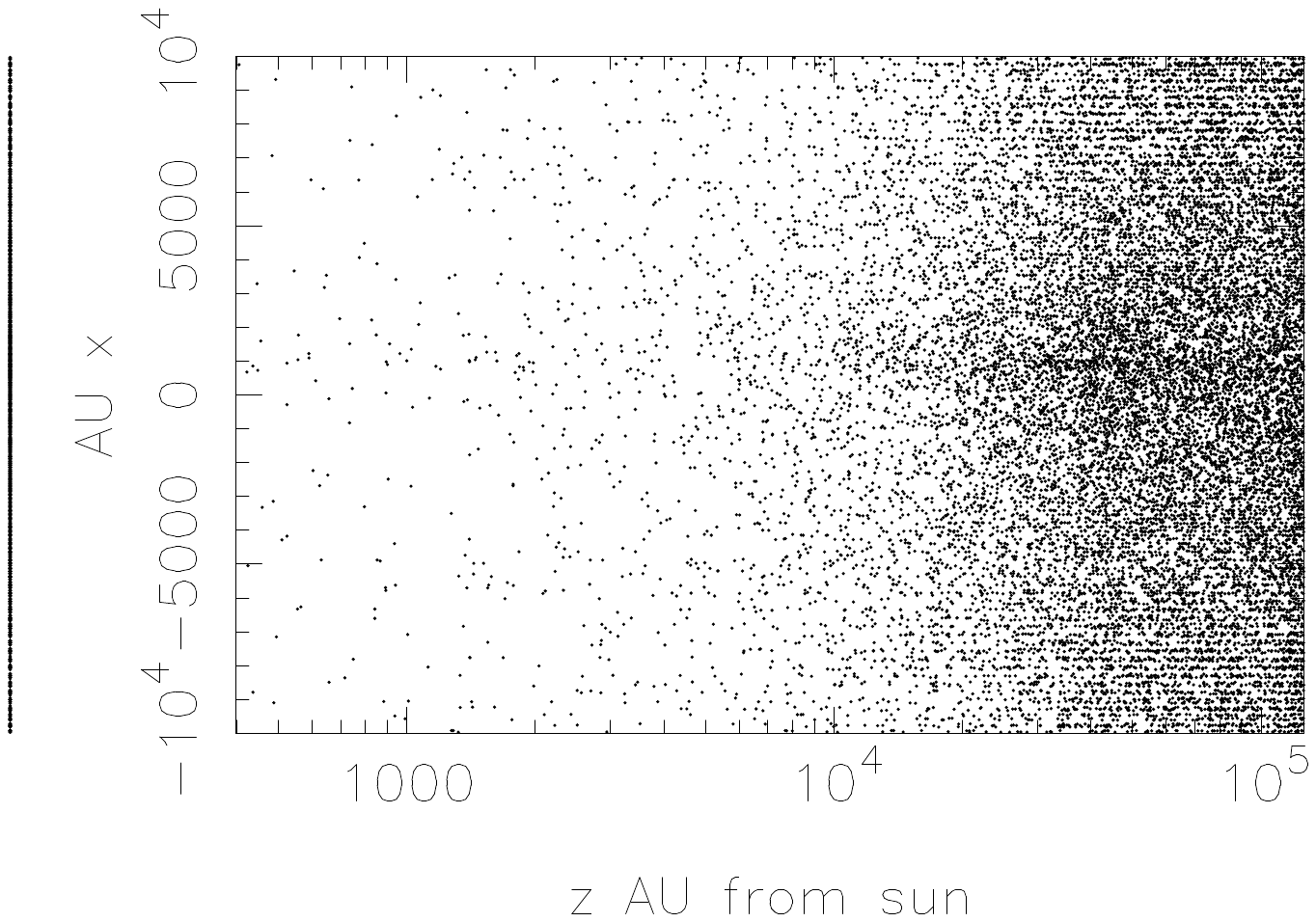}
\caption{Distribution of particles in the xz plane 0.05 Myrs into run 25.}
\end {figure}
\begin{figure}
\hspace{-0.5cm}
\includegraphics[width=0.7\textwidth]{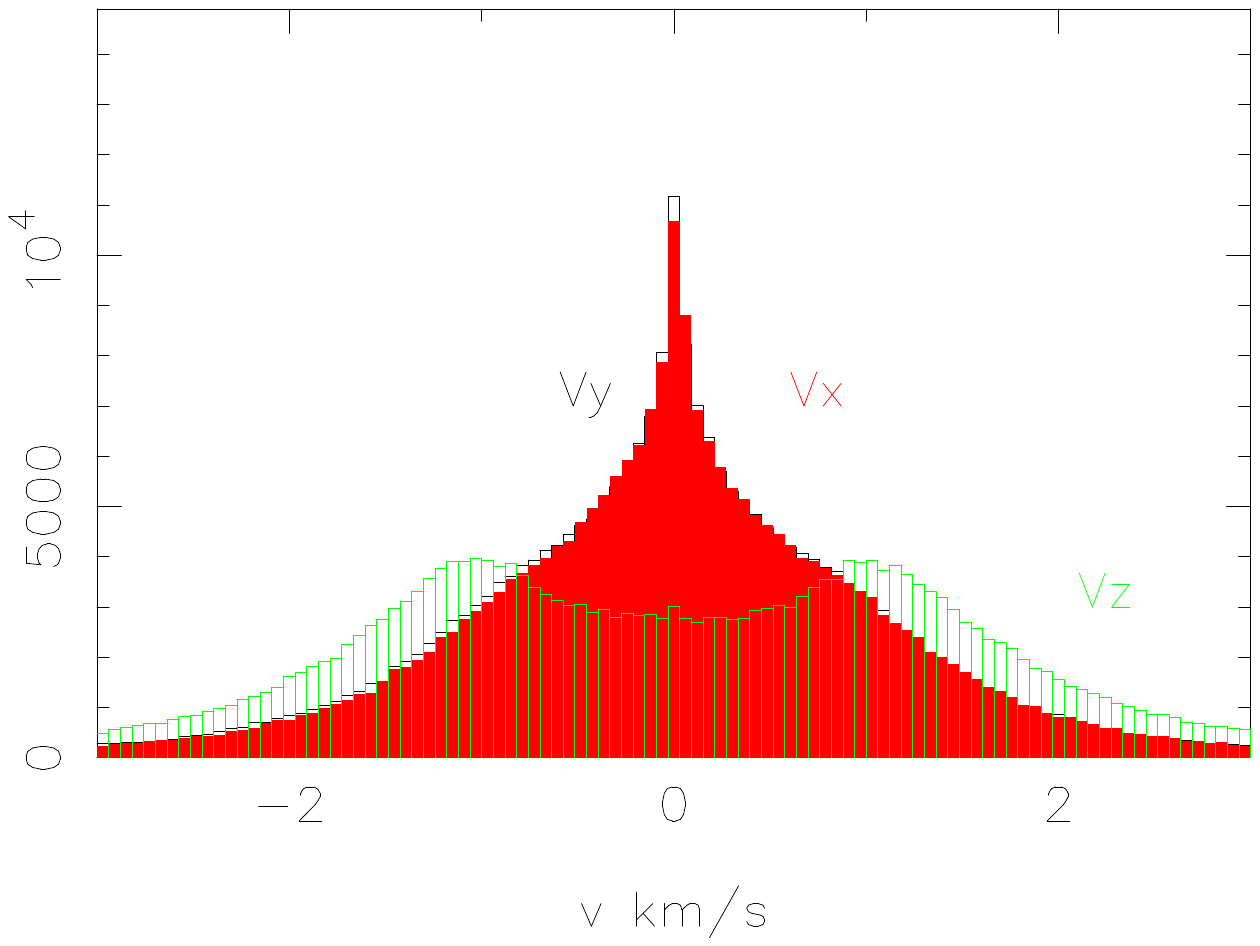}
\caption{Distribution of velocities 0.05 Myrs into run 25.}
\end {figure}

There are two factors for which the simulation comet finding rate needs
to be corrected, (1) the lost area factor in which DM particles
falling outside the maximum adopted impact parameter are neglected (4 $a^2/b_{max}^2$/N) and (2) the number of protocomets in the simulation, compared with N. 
Each of these two multipliers contributes a factor of N to the numerator in equation (3).
With these corrections
the predicted rate of entry of comets
	into the inner solar system\footnote{ 
The Oort cloud is generally considered a spherical or almost spherical region, 
	but it also has an inner, disc-shaped component that is more aligned with the ecliptic plane (the plane of the solar system). The outer Oort cloud is a vast, spherical halo extending far beyond the Kuiper Belt, while the inner Oort cloud is 
thought to be more of a donut shape and closer to the ecliptic. 
	} is
$${\rm rate}_2 = \frac{{\rm rate}_1~~\Delta t~N^2~ b_{max}^2~n}{4~a^2~ 2.5 \times 10^5
~\Delta t}
	= 1.6 \times 10^{-3}~n~ b_7^2~ {\rm rate}_1~,       \eqno(3)$$
where n is the number recorded in the simulation in time $\Delta$t,
	and b$_7$~=~b$_{max}~\times$~10$^7$. Figures~2~\&~3 show the particle distribution 
	and their velocity distribution shortly after the start of a simulation. 
    In Table I the number of comets observed in the inner solar system for each run is n. The maximum velocity attained by any of these is given in the centre column.
    It is clear that some comets, due to a decrease in orbital velocity, enter orbits of what are possibly observed as New comets. Depending on the direction of the primary black hole's velocity vector, other comets may escape the solar system due to an increase in orbital velocity. Note values as high as 80 \kms in Table 1. This is an important aspect of this problem, interesting in light of modern observations of interstellar comets.
\begin{table}[h]
\caption{Simulations}
	\vspace{1mm}
\begin{tabular}{llrrrrrrrr}
	\hline
	run&M&b$_{max}$&n&$\Delta$t&max v&$<\epsilon>$&Nb$_7$/a&rate$_1$&rate$_2$\\
	\#&M$_\odot$&AU&&Myrs&\kms&&&yr$^{-1}$&yr$^{-1}$\\
	\hline
	19&10$^{-9}$&10$^{-7}$&20&31.5&1.5&0.5&4000&62&2.0\\
	20&10$^{-8}$&10$^{-7}$& 7& 5.5&3.4&0.7&4000&6.2&0.07\\
	21&10$^{-7}$&10$^{-6}$& 7& 3.7&2.5&0.7&40000&0.62&0.7\\
	22&10$^{-8}$&10$^{-6}$&20&28.1&1.6&0.74&40000&6.2&20\\
	23&10$^{-9}$&10$^{-6}$&2 &18.1&1.4&1.14&40000&62&20\\
	24&10$^{-9}$&10$^{-6}$&0 &18.1&--&--&40000&62&0\\
	25&10$^{-6}$&10$^{-6}$&41&1.8&80.4&1.0&40000&0.062&0.41\\
	26&10$^{-9}$&10$^{-8}$& 7&5.2&4.5&0.79&400&62&0.07\\
	27&10$^{-8}$&10$^{-8}$&16 &5.2&45.2&1.62&400&6.2&0.0016  \\
	28&10$^{-10}$&10$^{-8}$&2  &5.5&1.8&0.62&400&616&0.02  \\
	29&10$^{-7}$&10$^{-8}$&25 &0.4&88.5&0.91&400&0.62&0.00025     \\
	30&10$^{-10}$&10$^{-9}$&3  &18.9&2.0&0.74&40&616&1.5$\times$10$^{-5}$   
 \\
	31&10$^{-10}$&10$^{-7}$&0  & 4.1& --& -- &40&616&0\\
	32&10$^{-6 }$&10$^{-5}$&3  &16.1&1.9&0.64&4$\times$10$^5$&0.062&3     \\
	33&10$^{-7}$ &10$^{-5}$&2  &8.1& 4.3 &2.2&4$\times$10$^5$&0.62&20\\
	34&10$^{-8}$ &10$^{-5}$&0  &16.9& -- &--&4$\times$10$^5$&6.2&0\\
	35&10$^{-5}$ &10$^{-5}$&43 &4.5&52.3&1.2&4$\times$10$^5$&6.2$\times$10$^
{-3}$&4.3\\
\hline
\multicolumn{10}{l}{$\epsilon$ is the ratio of $\mid$radial$\mid$ to tangential 
velocities; velocities are heliocentric.}\\
\multicolumn{10}{l}{rate$_1$ is PBH arrival; rate$_2$ is comet entry to the inner
	solar system.}\\
\end{tabular}
\end{table}
	\begin{figure}
		\hspace{-2 cm}		\includegraphics[width=0.7\textwidth]{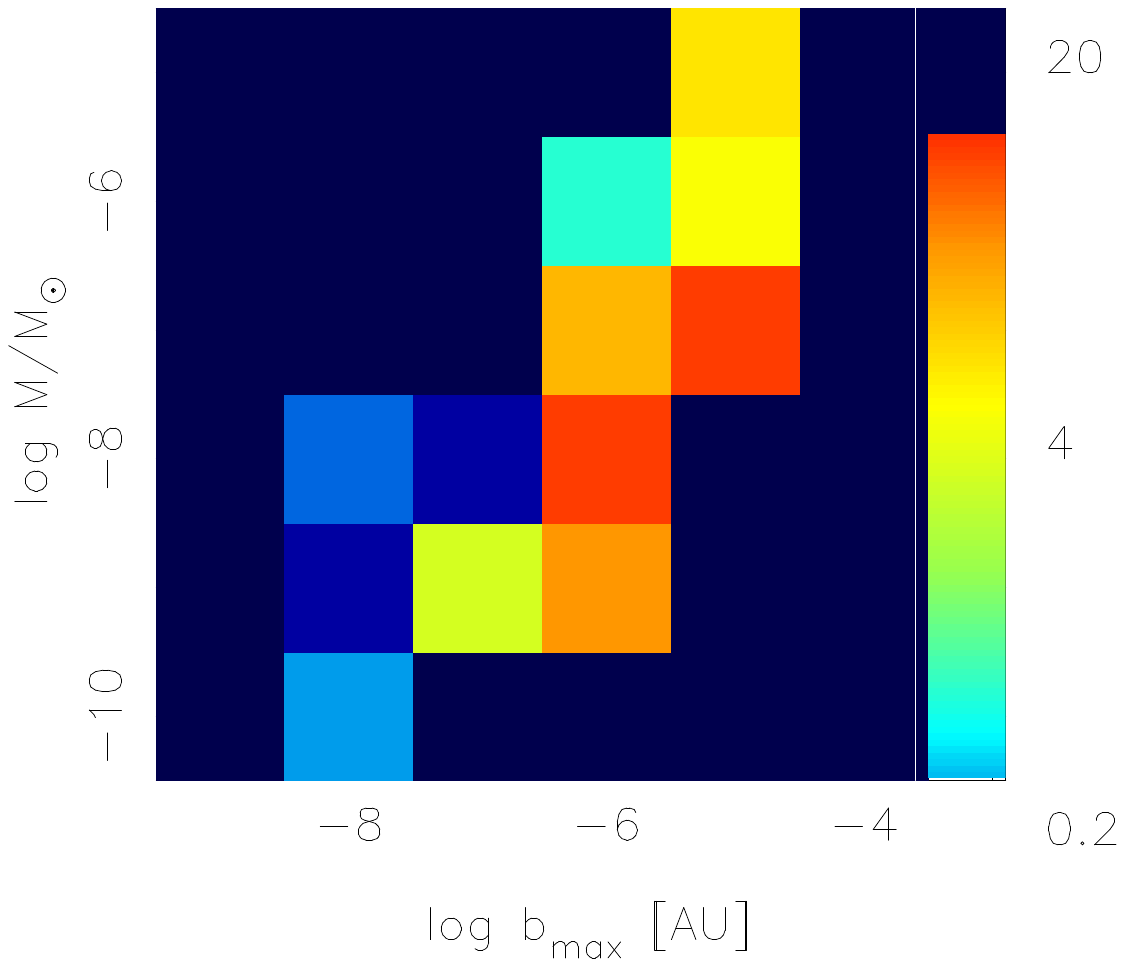}
		\caption{Simulated comet arrival annual rate in the inner solar system
		as a function of DM mass and maximum impact parameter.
		On the right is the colour scale which is logarithmic
		and ranges from 0.2 to 20. Deeper blue $<$ 0.1.}
	\end{figure}
	\subsection{Results}
	Figure 4 shows rate$_2$ for a range of DM masses and 
	maximum impact parameters, assuming that all the DM has mass M,
	i.e. f$_{PBH}$ = 1 in the case of PBH. The comet arrival rate
	in the inner solar system rate$_2~>$ 10 per year for some maximum
	impact parameter for 10$^{-9} <$ M $<$ 10$^{-5}$ M$_\odot$.
	The rate would be brought down to order unity for f$_{PBH}$ = 0.1. 
    Microlensing constraints on PBH in this mass range are given in Table II from Green (2026), Sugiyama \etal (2026) and Key (2025). In the upper part of the mass range in Table~1 a mass fraction larger than 0.1 is ruled out by these measurements. The OGLE results are more stringent, but anomalous with respect to the other experiments. The "asteroid window" (10$^{-9}$ M$_\odot$ and below) remains fully open because of diffraction limitations to microlensing.
    \input tablea1.tex

The shallower Oort cloud distributions shown in Figure 1 would be expected
to reduce the comet rate proportionally to the median radius. This was
borne out in re-runs of \# 19 and 20 with an r$^{-3.5}$ density dependence.
\section{Comet supply and demand}

An equally  fraught extrapolation to equation (3) in the
present work is the estimation of the number of comets inside 300 AU from the known number inside 10 AU.  
Integrating d($\frac{1}{2} v^2$)/dr = --GM/r$^2$, the infall equation in the gravitational field is

$$dr/dt = \surd 2(1/a - 1/r)^{1/2} = -\surd 2/r^{1/2 }~~~~ {\rm for~ small~ r}~~~\eqno(4)$$

\noindent in the adopted units of velocity (30 \kms) for which, after integration, the steady state radial distribution in spherical 
symmetry is enclosed number n $\sim$ r$^{3/2}$.

	Figure 5 from the JPL small bodies database\footnote{
	https://ssd.jpl.nasa.gov}	shows n = 500 inside 10 AU; so n 
$\approx$
82,000 inside 300 AU. With the rate found in the previous section, this can be furnished in a fraction of an Myr. Further orbital parameters are in Figure 6. But we need to consider the attrition rate.
\begin{figure}[H]
	\hspace{2 cm}
	\includegraphics[width=1.0\textwidth]{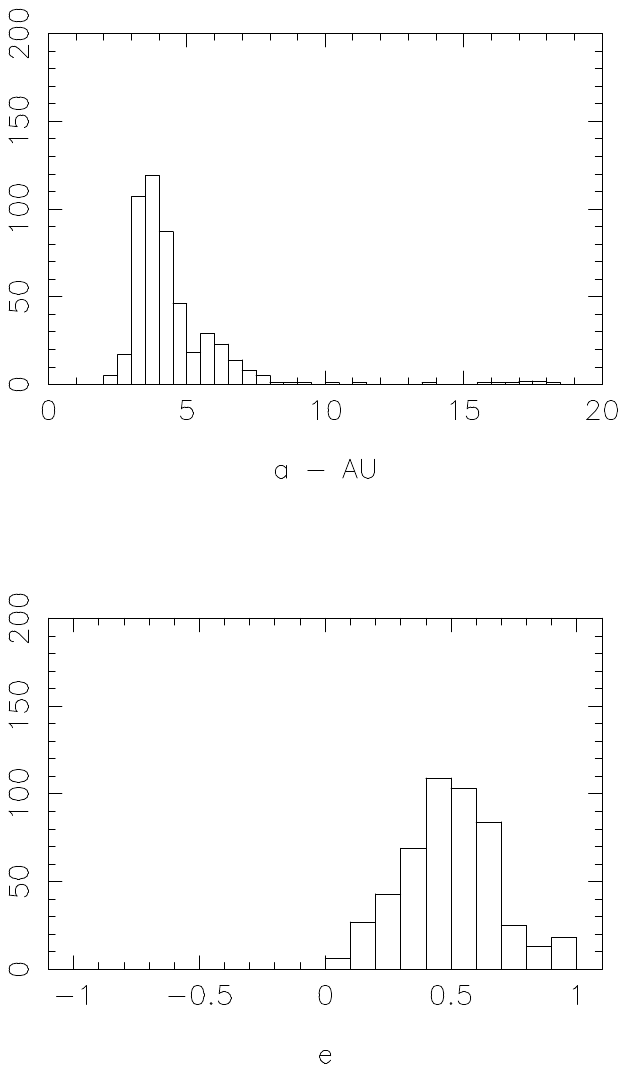}
	\caption{Numbered comet semi-major axes from the JPL small bodies
	database (above). Orbital eccentricities (below).}
	\includegraphics[width=1.0\textwidth]{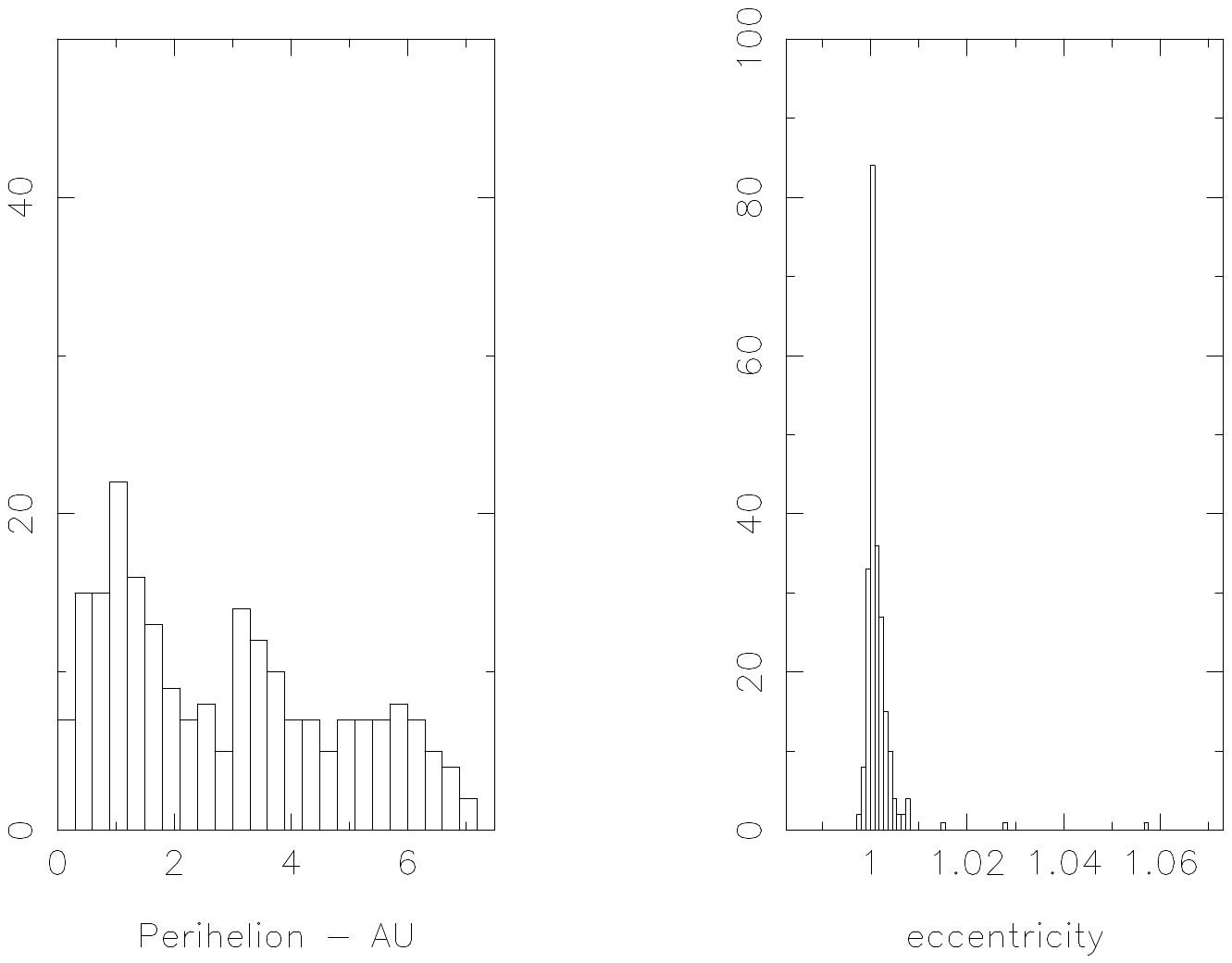}
	\caption{Perihelia and orbital eccentricities from Krolikowska \& Dybczynski (2020).}
\end{figure}

The COSINE project (Kwon \etal 2025) used NASA's WISE and NEOWISE mission to study 484 comets over 15 years. Of these 68 were on hyperbolic orbits (HC) and 140 
were on near parabolic (NPC) orbits . If this sample is typical of our 300 AU volume, 
and if we consider only the attrition from HCs, the rate is 1.5 per year.
If we add in, say, half the NPCs, the rate is double that, which might be hard to sustain
for f$_{PBH}~<$ 0.1.

	\section{Discussion}
	Measuring the fraction of the DM within the mass range we have trialled 
here that is PBH + FFM is well within the capabilities of the Rubin
	and $Roman$ telescopes. The optimum microlensing strategy is high cadence and moderate field. This is radically
	different from that adopted in Rubin's LSST (Large Survey of Space \& 
	Time), which is hemispheric and low
	cadence. If it is important to the stability of the Oort cloud, or for more general reasons, 
	a special DM survey might be commissioned (see Appendix). 
The cadence of $Roman$ is still being designed, but its aperture $\times$ field-
	of-view is 375 times less than Rubin's.
Microlensing has its limitations, of course, dilution by finite source effects and wave optics.
	These affect the limiting mass detectable, $\sim$ 10$^{-11}$ M$_\odot$. 
Moving to the ultraviolet
	can gain back a factor of 2; the source star counts contain more main sequence stars and the wave optics less
	of an issue.
	
	The toy model presented here has looked into one or two of the parameters
	affecting the arrival of comets in the region of the solar system where
	we can observe them. A further key parameter is comet mass dispersion.
	More sophisticated models exist (Bailey 1996), and these could  
	be adapted to further explore dark matter's involvement with comets.
	The sculpting of comets by the giant planets is an important factor omitted here.

     It should be noted that b$_{max}$ is, properly speaking, a gratuitous parameter. In a computing world where one could launch billions of particles into orbit from the Oort cloud, it would not be necessary to neglect incoming macroscopic dark matter with large impact parameter; they could be calculated explicitly. Future work should include not only major planets, but also asymptotically large values of b$_{max}$, or a maximum impact parameter equal to the mean separation between protocomets in the Oort cloud.
	\section{Conclusions}
	Depending on the fraction of DM that is of roughly lunar mass, DM
	may  partially (or perhaps, negligibly) contribute to the arrival of comets in the inner solar system.
	If $\sim$10\% of the DM is PBHs,
	not only may  DM be responsible for forming our Galaxy, but also DM
	(or FFM), 
	may have had a role (Biver \etal 2024) in forming our water world, a key
 requirement for life.
\section*{References}
\noindent Aalbers, J. \etal A next-generation liquid xenon observatory for dark matter and neutrino physics, 2023, JPhG, 50,  3001\\
Alcock, C. \etal Possible gravitational microlensing of a star in the Large Magellanic Cloud, 1993, Nature, 365, 621\\
Bailey, M. The Provenance and Evolution of Comets, 1996, EM\&P, 57, 72\\
Barberio, E. \etal The SABRE South technical design report executive summary,
2025, JInst, 20, 4001\\
Binney, J. Extracting science from surveys of our Galaxy, 2011, Prama, 77, 39\\
Binney, J. \& Vasiliev, E. Self-consistent models of our Galaxy, 2023 MNRAS, 520. 1832\\
Biver, N.,  Dello Russo, N. ,  Opitom, C. \&  Rubin, M., Chemistry of Comet Atmospheres
	2024, {\it Comets III}, Ed.  Karen  Meech \etal Space Science Series, UA
 Press,  pp. 459-498\\
 Brasser, R., Duncan, M. \& Levison, H.  Embedded star clusters and the formation of the Oort Cloud, 2006, Icarus, 184, 59\\
Brown, G., He, L. \& Unwin, J. The Potential Impact of Primordial Black Holes on Exoplanet Systems, 2025,
OJA, 8, 162\\
Caldwell, J. \& Ostriker, J. The mass distribution within our Galaxy - A three component model, 1981, ApJ,  251,  61\\
De Rocco, W., Ganymede's subsurface ocean as a dark matter detector, 2025, PRD, 112, 5023\\
de Salas, P. \& Widmark, A. Dark matter local density determination: recent observations and future prospects, 2021, RPP, 84, 4901\\ 
Dehnen W., A Family of Potential-Density Pairs for Spherical Galaxies and Bulges, 1993, MNRAS, 265, 250\\
Dones, L., Weissman, P., Levison, H. \& Duncan, M., Oort cloud formation and dynamics, 2004, in $Comets~II$,
eds. M. Feston, H. Keller \& H. Weaver (Tucson: UA Press), p.153\\
Duncan, M., Quinn, T. \& Tremaine, S. The Formation and Extent of the Solar System Comet Cloud, 1987, AJ, 94, 1330\\ 
Fairbairn, M., Marsh, D. \& Quevillon, J. Structure formation and microlensing with axion miniclusters 2018, PRL, 97, 3502\\
Freeman, K. The old population, 1985, IAUS, 106, 113\\
Green, A., Stellar microlensing as a probe of Primordial Black Holes: status and prospects, 2026, arxiv 2602.15974\\
Harrington, R. Implications of the observed distributions of very long period comet orbits,
1985, Icarus, 61, 60\\
Heisler, J. \& Tremaine, S., The influence of the Galactic tidal field on the Oort comet cloud 1986, Icarus, 65, 13\\
Higuchi, A., Kokubo, E., Kinoshita, H. \& Mukai, T. 2007, Sedna and the Oort Cloud around a migrating Sun, AJ, 134, 1693\\
Kaib, N. \& Quinn, T. The formation of the Oort cloud in open cluster environments, 2008, Icarus, 197, 221\\
Kaib, N.,  Roskar, . \& Quinn, T., Sedna and the Oort Cloud around a migrating Sun, 2011, Icarus, 215, 491\\
Key, R. Determining the Density of Primordial Black Holes as Dark Matter via Microlensing 2025, PhD thesis, Swinburne University\\
Krolikowska, M. \& Dybczynski, P., The catalogue of cometary orbits and their dynamical evolution, 2020, A\&A, 640, 97\\
K\"uhnel, F. 2025, "Positive Indications for Primordial Black Holes",
{\it Primordial Black Holes}, ed Christian Byrnes, Springer, p.453\\
Kwon, Y. \etal COSINE (Cometary Object Study Investigating Their Nature and Evolution). I. Project Overview and General Characteristics of Detected Comets, 2025, ApJS, 280, 67\\
Murai, K., Sakurai, K. \& Takahashi, F. Primordial black hole formation via inverted bubble collapse 2025, JHEP, 7, 65, \\
Nesvorn\'y D., Dynamical Evolution of the Early Solar System 2018, ARAA, 56, 137\\
Planck collaboration, Planck 2018 results. VI. Cosmological parameters, 2020, A\&A, 641, A6\\
Ortiz, R. \& Lepine, J. A model of the galaxy for predicting star counts in the infrared, 1994, A\&A, 279, 90\\
Romao, M., Croon, D., Crossey, B. \& Godines, D., Dark classification matters: searching for primordial black holes with LSST, 2025, MNRAS, 543, 351\\
Rickman H. Stellar Perturbations of Orbits of Long-period Comets and their Significance for Cometary Capture, Bull. Astron. Inst. Czechoslovakia 1976. V. 27, 92.\\
Stothers, R. Mass extinctions and missing matter, 1984, Nature, 311, 17\\
Sugiyama, S. \etal Microlensing constraints on Primordial Black Hole abundance with Subaru Hyper Suprime-Cam observations of Andromeda, 2026, arxiv 2602.05840\\
Vida, D. \etal Direct measurement of decimetre-sized rocky material in the Oort cloud, 2023, Nat As, 7, 318\\
Wajer, P. \etal , Oort Cloud and sednoid formation in an embedded cluster, I: Populations and size distributions 2024, Icarus, 415, 116065\\
\section*{Declarations}
\begin{itemize}
\item Funding: I acknowledge ARC grant CE200100008 which, together with the five Australian university nodes, funds Centre for Dark Matter Particle Physics research.
\item Data availability: Data are available from the author. 
\item Code availability 
The code comet6.f is available on github: 
github.com/jrmould/darkmatter, together with a README file.
\end{itemize}
\appendix
\setcounter{figure}{0}
\renewcommand{\thefigure}{A\arabic{figure}}
\section*{Appendix}
The  PINGAS Galactic star count model 
(Lepine \& Ortiz 1994)
without obscuration, together with the Binney \& Vasiliev (2023) mass  model of 
the Galaxy, was used to
predict the microlensing event rate under the assumption that all the DM was lunar mass PBHs and
the simplifying assumption that stars more distant than the Galactic centre were
 sources.
Figure A1 is the result. An optimum cadence for microlensing was assumed, e.g. 
time before return-to-field = 15 minutes.
The comparison with Figure~5 of Romao \etal (2025) (LSST's actual cadence)
is informative.
\begin{figure*}
	\includegraphics[width=1.2\textwidth]{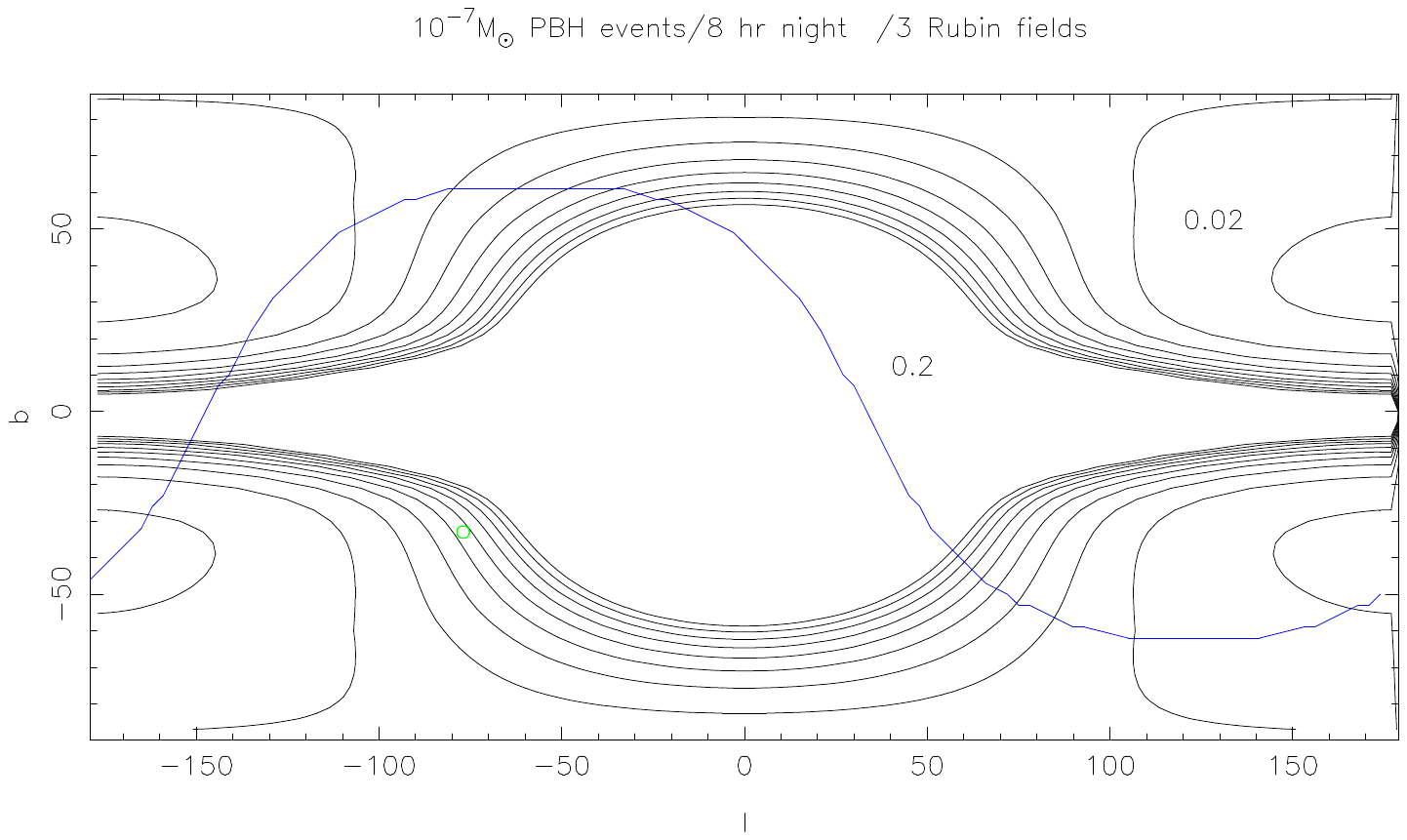}
	\caption{Microlensing event rate per 8 hour night for 3 Rubin fields for
 100\% lunar mass DM
	and an optimum cadence for detecting and mapping them.
	The lowest contour is 0.02 and the highest 0.2. The blue curve is the celestial equator.
	The green dot is the Large Magellanic Cloud, microlensing target for the
 MACHO project
	(Alcock \etal 1993) and others. Only Galactic stars enter the calculation as source stars;
	so an uptick for the LMC is absent. Compare this with the anticipated results from
	the LSST cadence, figure 5 of Romao \etal (2025.)}
\end{figure*}

\section*{Acknowledgements}
Figure 5 is based on JPL's small bodies database,  sponsored by NASA under Contract NAS7-030010. JPL is operated by Caltech for NASA.  
I acknowledge use of the
Swinburne University's Ozstar \& Ngarrgu Tindebeek supercomputers, the latter named by Wurundjeri elders and translating as "Knowledge of the Void" in the local Woiwurrung language. I thank  colleagues in our microlensing team and the CDMPP for helpful discussions and the referees for improvements to the paper.
The PINGAS star counts code is at http://www.astro.iag.usp.br/$\sim$jacques/pingas.html.



\end{document}

%% file: tablea1.tex
\begin{table}
	\caption{Microlensing limits on PBH mass fraction}
\begin{tabular}{lrrrrrr}
	\hline
Survey &Duration &FoV &\# stars&Detected&Mass range&f$_{PBH}$\\

        & (years)&(deg$^2$)&(million)&\# events&M$_\odot$&   \\
\hline
MACHO &5.7 &14& 12 &13& 3 $\times$ 10$^{-7}$ to 5 $\times$ 10$^{-4}$ &f $\leq$ 0.1\\
EROS&$\sim$12 &88* &25*& 1& 10$^{-6}$ to 0.01& f $<$ 0.1\\ 
OGLE&$\sim$20 &105$^\dagger$ & 78.7$^\dagger$ & 13&  10$^{-8}$ to 0.01 &f $<$ 0.01\\
Subaru HSC&39.3 hrs&1.7&80&4&10$^{-7}$ to 10$^{-6}$ &0.01 to 0.1\\

	AMDC&5 nights&3&4&1&10$^{-8}$ to 10$^{-5}$&f $\sim$ 0.1\\
\hline
\multicolumn{7}{l}{*EROS-2,~~~$^\dagger$OGLE-IV~~~Data from Green (2026)
	~~~FoV = field of view}\\
\end{tabular}
\end{table}